\documentstyle[manuscript,eqsecnum,aps,version2]{revtex}
\draft

\def\Ds{D \!\!\!\! / \,}

\begin{document}

\title
\bf Chiral Symmetry Restoration and $Z_N$ Symmetry
\endtitle
\author{Peter~N.~Meisinger and Michael~C.~Ogilvie}
\instit
Department of Physics, Washington University, St. Louis, MO 63130
\endinstit
\medskip
\centerline{\today}

\abstract
We demonstrate that chiral symmetry restoration in quenched finite temperature
QCD depends crucially on the $Z_3$ phase of the Polyakov loop ${\cal P}$.
This dependence is a general consequence of the coupling
of the chiral order parameter to the Polyakov loop.
We construct a model for chiral symmetry breaking and restoration which includes
the effect of a nontrivial Polyakov loop by calculating the effective potential
for the chiral condensate of a Nambu-Jona-Lasinio model in a uniform 
temperature dependent $A_0$ gauge field background. Above the deconfinement
temperature there are three possible phases corresponding to the $Z_3$
symmetric phases of the Polyakov loop in the pure gauge theory. In the phase in
which ${\rm tr_c}({\cal P})$ is real and positive the first order deconfining
transition induces chiral symmetry restoration in agreement with simulation
results. In the two phases where $Re[{\rm tr_c}({\cal P})] < 0$ the sign of
the leading finite temperature correction to the effective potential is
reversed from the normal phase, and chiral symmetry is not restored at the
deconfinement transition; this agrees with the recent simulation studies of
Chandrasekharan and Christ. In the case of $SU(N)$ a rich set of
possibilites emerges. The generality of the mechanism makes it likely
to occur in full QCD as well; this
will increase the lifetimes of metastable $Z_3$ phases.

\endabstract


\pagebreak[4]

\section{INTRODUCTION}
\label{s1}

Restoration of chiral symmetry is observed at high temperatures in simulations
of both quenched and full QCD. In interesting recent work Chandrasekharan and
Christ have shown through simulation that the chiral properties of quenched
finite temperature QCD depend strongly on the $Z_3$ phase of the Polyakov loop
${\cal P}$ \cite{ChCh}. Only in the phase where $Re[{\rm tr_c}({\cal P})] > 0$
and $Im[{\rm tr_c}({\cal P})] = 0$ is chiral symmetry restored at the
deconfinement transition. In the other two $Z_3$ phases, chiral symmetry is not
restored. In this letter we provide a theoretical explanation for these results
and make predictions for the general case of $SU(N)$.

The physical mechanism which couples the chiral condensate to the Polyakov loop
is simple: finite temperature corrections to the quark determinant involve
quark trajectories which are topologically non-trivial in the Euclidean
temporal direction. Such trajectories are naturally weighted by powers of the
Polyakov loop. When $Re[{\rm tr_c}({\cal P})] $ changes sign, the sign of the
leading finite temperature correction to the effective potential for the chiral
condensate changes as well.

In our efforts to understand the results of Chandrasekharan and Christ, we have
studied an effective $U(N_f) \times U(N_c)$ chirally invariant
Nambu-Jona-Lasinio (NJL) model coupled to a uniform temperature dependent
$A_0$ gauge field. NJL
models have been analyzed in great detail, including their finite temperature
properties \cite{HaKu,Kl}. We extend these
models by allowing a dependence on the uniform external gauge field $A_0$.
The $A_0$ influence on the phase structure of this model manifests
itself solely through the Polyakov loop. The Polyakov loop is taken to be
identically zero for temperatures below the deconfinement temperature $T_d$ and
jumps to a nonzero value at $T_d$,
as in pure $SU(3)$ gauge theory. The deconfining phase transition is known
to be first order for SU(3) and believed to be first order for all SU(N) with
$N > 3$; SU(2) exhibits a second order phase transition and must be considered
separately.

We use the effective potential formalism to study our extended model. Since
closed quark loop effects are by definition absent in quenched QCD, it may at
first glance seem strange to consider an effective potential which contains a
contribution from the quark determinant. However, consider diagram (a) in
Fig.~\ref{f1}. It is one of an entire class of diagrams which presumably
contribute to chiral symmetry breaking in the quenched approximation.
It has long been argued that such
diagrams are treated approximately in NJL models by the interaction shown in
diagram (b) \cite{Ko}. If we regard
the effective potential as simply a tool to derive the more fundamental gap
equation for the chiral condensate, then the potential must include 
terms such as the one shown in diagram (d) of Fig.~\ref{f1}, since they are
obtained from Fierz transformations of quenched QCD terms like the one shown
in diagram (c).

\section{EFFECTIVE POTENTIAL}
\label{s2}

We begin with a Euclidean action of the form
\begin{eqnarray}
{\cal L}_E = \bar{\Psi}(\Ds + m) \Psi
- G \sum_{a = 0}^{N_f^2 - 1} \left[ (\bar{\Psi} \lambda^a \Psi )^2 +
(\bar{\Psi} i \gamma_5 \lambda^a \Psi )^2 \right]
\label{e1.1}
\end{eqnarray}
where the quark field $\Psi$ carries flavor and color \cite{Kl}.
The $\lambda^a$ are the generators of $U(N_f)$.
The covariant derivative
reflects the presence of a uniform background temporal $SU(N_c)$ gauge field:
\begin{eqnarray}
D_{\mu} = \partial_{\mu} + igA_{\mu}(x) \quad {\rm with} \quad
A_{\mu}(x) = \delta_{0\mu}A_0.
\label{e1.2}
\end{eqnarray}

Let $S_0$ denote the propagator for a free quark with mass $m$, and let $S$
denote the quark propagator of the interacting theory. Following Cornwall,
Jackiw, and Tomboulis \cite{CoJaTo}, an effective action $\Gamma(S)$ for the
theory in Eq.~(\ref{e1.1}) is given by
\begin{eqnarray}
\Gamma(S) = {\rm Tr} \left[ \ln \left( S_0^{-1} S \right) \right]
- {\rm Tr} \left( S_0^{-1} S  - 1 \right) - G \sum ({\rm diagrams}),
\label{e2.1}
\end{eqnarray}
where the diagrams are shown in Fig.~\ref{f2} with each internal fermion line
being associated with the propagator $S$. The two rightmost
diagrams sum to zero if $S(x,x)$ is a scalar in spinor space. In the presence
of a uniform background $A_0$ gauge field this scalar property does not hold at
finite temperature. For simplicity,
however, we consider only the first diagram. This is
the Hartree approximation and the leading term in a $1/N_c$ expansion.
It follows that
\begin{eqnarray}
\Gamma(S) = {\rm Tr} \left[ \ln \left( S_0^{-1} S \right) \right]
- {\rm Tr} \left( S_0^{-1} S  - 1 \right) - 2 G \int_0^\beta dt \int d^3 \vec{x}
\, \sum_f \left\{ {\rm tr_{cd}} \left[ S_f(x,x) \right] \right\}^2 
\label{e2.5}
\end{eqnarray}
where the sum is over all flavors and the subscripted trace is over Dirac and
color indices alone. $\beta$ is the inverse temperature. The
equation for the propagator is
\begin{eqnarray}
S_f^{-1}(x, y) = S_{0f}^{-1}(x, y) + 4 G  {\rm tr_{cd}}\left[ S_f(x,x) \right]
\delta(x, y).
\end{eqnarray}
Letting $\sigma_f = {\rm tr_{cd}}\left[ S_f(x,x) \right]$, the quark mass of a
single flavor in the interacting theory is given by $M_f = m_f + 4 G \sigma_f$.
Conventionally, $m_f$ and $M_f$ are considered to be the current and
constituent quark masses, respectively.

The phase structure of the model is conveniently studied using the effective
potential
\begin{eqnarray}
V(\sigma, A_0, T) = \sum_f \Biggl\{ 2 G \sigma_f^2
- 2 \sum_{k_0} {1 \over \beta} \int {d^3 \vec{k} \over (2 \pi)^3}
{\rm tr_c} \ln \left[ (k_0 + gA_0)^2 + \omega_{f \vec{k}}^2 \right] \Biggr\},
\label{e2.10}
\end{eqnarray}
where
$\omega_{f \vec{k}} = [ \vec{k}^2 + (m_f + 4 G \sigma_f)^2 ]^{1/2}$,
and the sum over $k_0$ is the mode sum over Matsubara frequencies. Evaluating
this mode sum and discarding an irrelevant constant,
\begin{eqnarray}
V(\sigma, A_0, T) &=& \sum_f \Biggl[ 2 G \sigma_f^2
-  2 N_c \int {d^3 \vec{k} \over (2 \pi)^3} \omega_{f \vec{k}} \nonumber\\
&\phantom{=}& \phantom{N_f \Biggl[}
- {2 \over \beta} \sum_{n = 1}^{\infty} {(-1)^n \over n}
{\rm tr_c} \left( {\cal P}^n + {\cal P}^{\dagger n} \right)
\int {d^3 \vec{k} \over (2 \pi)^3} e^{- n \beta \omega_{f \vec{k}}} \Biggr],
\label{e2.12}
\end{eqnarray}
where ${\cal P} = \exp (i \beta g A_0)$
is the Polyakov loop associated with the background field $A_0$. The second
term in Eq.~(\ref{e2.12}) is an integral representation of the zero-temperature
quark determinant which sums all possible zero-point energies. The infinite
series of integrals in the third term represents the finite temperature
corrections. Intuitively, each order in $n$ is associated with quark paths that
wind around space-time in the temporal direction a net number of times $n$.
This is essentially an image expansion of the finite temperature quark
determinant \cite{MeOgI,MeOgII}. The standard NJL model is recovered by setting
$A_0$ to zero.

The evaluation of the quark determinant at finite temperature is slightly
subtle. After regularization, the zero-temperature portion of the determinant
contains an $M^4 \log (M)$ term. This logarithmic term is cancelled by a
similar term from the finite temperature contribution; every order in the
summation over $n$ contributes to this cancellation. On the other hand, the
integrals associated with each order in $n$ can be performed analytically, with
no need for the imposition of a cutoff, resulting in modified Bessel functions
$K_n$. It is tempting to approximate the finite temperature correction by simply
keeping the first few terms in the series. However, this approximation fails to
reproduce the correct behavior of the effective potential for small constituent
masses and can lead to unphysical behavior.

The image expansion nevertheless remains an effective tool for understanding
the role of $Z_N$ phases in critical behavior. The basic physics for the change
in critical behavior associated with the Polyakov loop is given by the $n=1$
term in Eq.~(\ref{e2.12}). Terms higher order in $n$ are suppressed by powers
of ${\rm exp}(-M/T)$, making this a reliable approximation when $T << M$.
Keeping only the $n=1$ term, the effective potential can be written as
\begin{eqnarray}
V(\sigma, A_0, T) &=& \sum_f \Biggl\{ 2 G \sigma_f^2
-  2 N_c \int {d^3 \vec{k} \over (2 \pi)^3} \omega_{f \vec{k}}
\nonumber\\
\smallskip \nonumber\\
&\phantom{=}& \phantom{\sum_f \Biggl\{}
+ \left[ {(m_f + 4 G \sigma_f) T \over \pi} \right]^2
K_2 \left( {m_f + 4 G \sigma_f \over T} \right)
{\rm tr_c} \left( {\cal P} + {\cal P}^{\dagger} \right) \Biggr\}.
\label{e2.14}
\end{eqnarray}
The crucial point is that the sign of the $n = 1$ finite temperature correction
to $V(\sigma, A_0, T)$ changes with the sign of
${\rm tr_c}({\cal P} + {\cal P}^{\dagger})$.
If the effect of finite temperature corrections is to raise the normal minimum
relative to $M = 0$, phases with $Re[{\rm tr_c}({\cal P})] < 0$ will instead
lower this minimum. This suppresses chiral symmetry restoration in these phases.

A quantitative demonstation is best performed with the full effective
potential. It is convenient to introduce a further approximation to powers of
the Polyakov loop:
\begin{eqnarray}
{\rm tr_c}({\cal P}^n) \approx N_c \left( {{\rm tr_c} {\cal P} \over N_c}
\right)^n.
\label{e2.15}
\end{eqnarray}
This approximation neglects the formation of triality zero states in the gluon
plasma. That is, each quark moves independently in its own fixed background. We
also regulate $V(\sigma, A_0, T)$ by introducing a non-covariant cut-off,
$\Lambda$. Denoting the trace of the Polyakov loop by
$\phi$, Eq.~(\ref{e2.12}) now can be resummed as
\begin{eqnarray}
V(\sigma, \phi, T) &=& \sum_f \Biggl\{ 2 G \sigma_f^2
- {N_c \over \pi^2} \int_0^{\Lambda} dk \, k^2 \omega_{fk}
\nonumber
\smallskip \nonumber\\
&\phantom{=}& \phantom{N_f \Biggl\{}
+ {N_c \over \pi^2 \beta} \int_0^{\Lambda} dk \, k^2
\ln \left[ 1 + e^{- \beta \omega_{fk}} \left( {\phi + \phi^* \over N_c} \right)
+ e^{- 2 \beta \omega_{fk}}
\left( {\phi^* \phi \over N_c^2} \right)
\right] \Biggr\}.
\label{e2.16a}
\end{eqnarray}
In this approximation, if $\phi$ is zero, all finite temperature effects are
absent from the potential.

Other approximations are also defensible. Consider
\begin{eqnarray}
V(\sigma, \phi, T) &=& \sum_f \Biggl\{ 2 G \sigma_f^2
- {N_c \over \pi^2} \int_0^{\Lambda} dk \, k^2 \omega_{fk} \nonumber
\smallskip \nonumber\\
&\phantom{=}& \phantom{N_f \Biggl\{}
+ {N_c \over \pi^2 \beta} \int_0^{\Lambda} dk \, k^2
\ln \left[ 1 + e^{- \beta \omega_{fk}} \left( {\phi + \phi^* \over N_c} \right)
+ e^{- 2 \beta \omega_{fk}}
\right] \Biggr\},
\label{e2.16b}
\end{eqnarray}
which differs from the previous expression only in the last term, where the
Polyakov loop for the quark is cancelled against its inverse for the antiquark.
In this approximation, some finite temperature effects of quark-antiquark pairs
are taken into account. In fact, even when $\phi = 0$, there is a chiral
symmetry restoration transition at exactly twice the critical temperature of
the ordinary NJL model.

\section{DISCUSSION AND NUMERICAL RESULTS}
\label{s3}

Following Ref.~\cite{Be}, we determine the noncovariant cut-off $\Lambda$ and
the constituent quark mass $M$ by fixing the chiral condensate and the pion
decay constant at zero temperature: $\sigma = (246.7 \,\, MeV)^3$ and
$f_{\pi}^2 = (93 \,\, MeV)^2$.  For the remainder of this letter we will use
the numerical solution $M = 335 \,\, MeV$, $\Lambda = 631 \,\, MeV$, and
$G \Lambda^2 = 2.2$, which assumes a current quark mass $m = 4 \,\, MeV$ and
$N_f = 2$ \cite{HaKu,Be,BeMeZa,MeArGo}.

The standard NJL model is recovered by setting $\phi = N_c$ in
Eq.~(\ref{e2.16a}). Chiral symmetry is restored by a second-order phase
transition; with the $\Lambda$ and $G$ given above, the critical temperature is
$T_c = 194.6 \,\, MeV$. Our
model will behave as the zero-temperature solution of the standard NJL model
until the Polyakov loop develops an expectation value; in particular, chiral
symmetry will remain broken while $\phi = 0$. Thus, in our model the critical
temperature for chiral symmetry restoration $T_c$ will always be greater than
or equal to the temperature for deconfinement $T_d$. We have shown recently in
Ref.~\cite{MeOgIII} that the unquenched version of this model has $T_c$
strictly greater than $T_d$ for certain ranges of phenomenological parameters.
However, in lattice simulations it appears that $T_c = T_d$. Requiring this
behavior places a mild constraint on this model as shown in Fig.~\ref{f3}.
At this time, the constraint is not a strong test of the model, because
$\phi$ is not known: the Polyakov loop is multiplicatively renormalized in a
way that depends on the regularization scheme. In particular, lattice
simulations do not directly measure $\phi$.
	
If the Polyakov loop is in one of the $Z_3$ phases of QCD in which
$\phi + \phi^* < 0$, the potential $V(\sigma_f, \phi, T)$ will not have a
minimum at $\sigma_f = 0$, unless the temperature is significantly greater than
the $T_c$ of the corresponding real phase. This is precisely the behavior
observed in quenched QCD simulations \cite{ChCh}.
In Fig.~\ref{f4} we compare the effective potentials of the real and complex
$Z_3$ phases for $\phi = 2.0$ and $T = 252.1 \,\, MeV$ to the
zero-temperature effective potential where $\phi = 0$. The relative depth of
the well in the complex phase has increased as expected
given the change in sign of $\phi + \phi^*$.

The general case of a theory with $SU(N)$ gauge fields can also be considered.
The mechanism will be the same whether the full theory or a
phenomenological, low energy approximation is studied. The sign of the most
important
finite temperature correction to the effective potential depends on the sign of
the real part of $\phi$. In general, the temperature dependence of the chiral
condensate will be different in different $Z_N $ phases. Only phases in which
the Polyakov loops are related by complex conjugation will have the same
behavior. Thus, for both $SU(2N)$ and $SU(2N + 1)$, there will be $N + 1$
distinct behaviors.

The finite temperature behavior when $\phi_d$ is purely imaginary is determined
by higher order terms in the image expansion [see Eq.~(\ref{e2.12})].
In $SU(4)$, scaling the
temperature by a factor of two yields a potential
identical to the original NJL model, except the remaining exponential in
Eq.~(\ref{e2.16a})
is multiplied by a factor of $|\phi_d|^2 / N_c^2$. Naively, this would imply
that the critical temperature is more than double that of the phase
where $\phi$ is real and positive.

In the high temperature phase of full QCD, the $Z_N$ phases are not equivalent;
quark effects lift the $Z_N$ degeneracy and only the phase in which the
Polyakov loop is real and positive is stable. The other phases are metastable,
at least in the standard Euclidean formalism. 
The action of the bounce solution has been calculated under the assumption that
chiral symmetry has been restored in all phases \cite{DiOg}. However, it seems
likely that the same mechanism discussed here for the quenched approximation is
also operative in the full theory. Because the strength of the $Z_N$ symmetry
breaking decreases with increasing mass, the absence of chiral symmetry
breaking in those phases will tend to decrease the action of the bounce
solution, and therefore increase the lifetime of the metastable states.
This will also modify the thermodynamic properties of these metastable
phases.

\section*{ACKNOWLEDGEMENTS}
We wish to thank the U.S. Department of Energy for financial support under
grant number DE-FG02-91-ER40628.

\figure{ Feynman diagrams contributing to the constituent quark mass
in quenched QCD (a) and the NJL model (b) and related diagrams
contributing to the effective potential (c and d, respectively).
\label{f1} }

\figure{ Feynman diagrams contributing to the effective action.
\label{f2} }

\figure{ Boundary in the ($\phi_d$, $T_d$)-plane between parameter values
which yield one phase transition and parameter values which yield two. Only
points above the curve are compatible with simulation results.
\label{f3} }

\figure{ Comparison of the effective potentials of the real and complex $Z_3$
phases for $\phi = 2.0$ and $T = T_{\chi}(\phi) = 252.1 \,\, MeV$ to the
zero-temperature effective potential where $\phi = 0$.
\label{f4} }

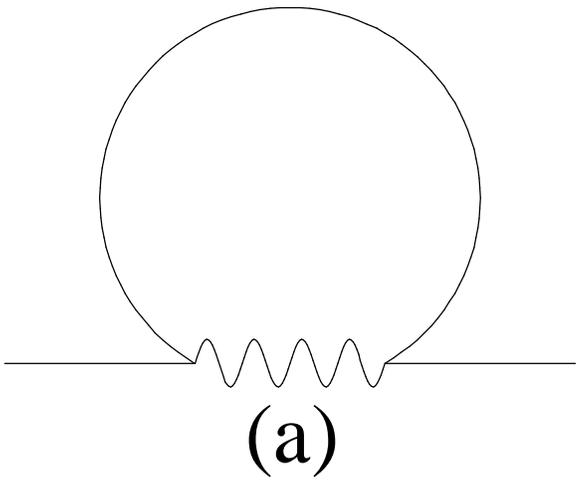
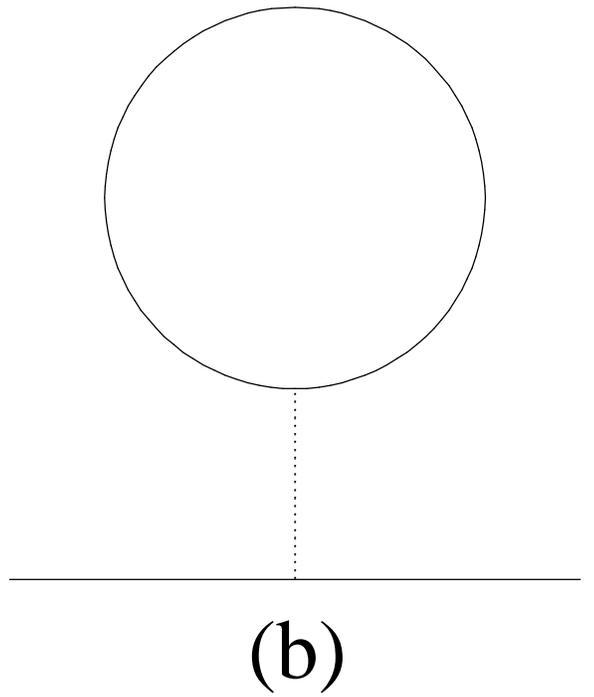
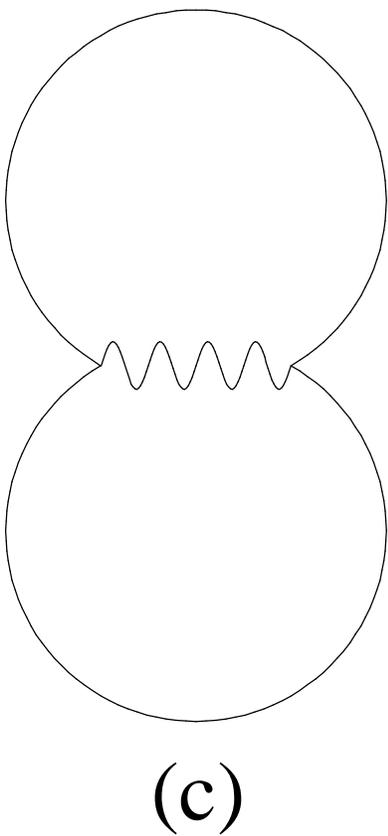
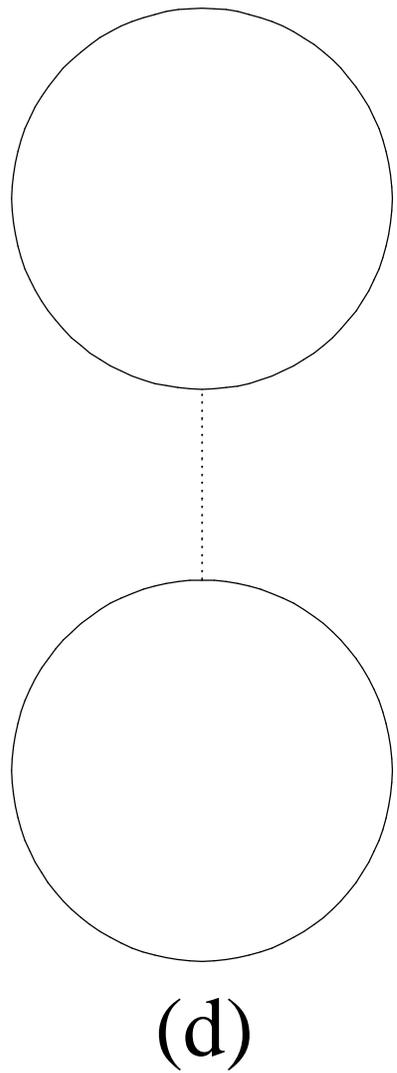
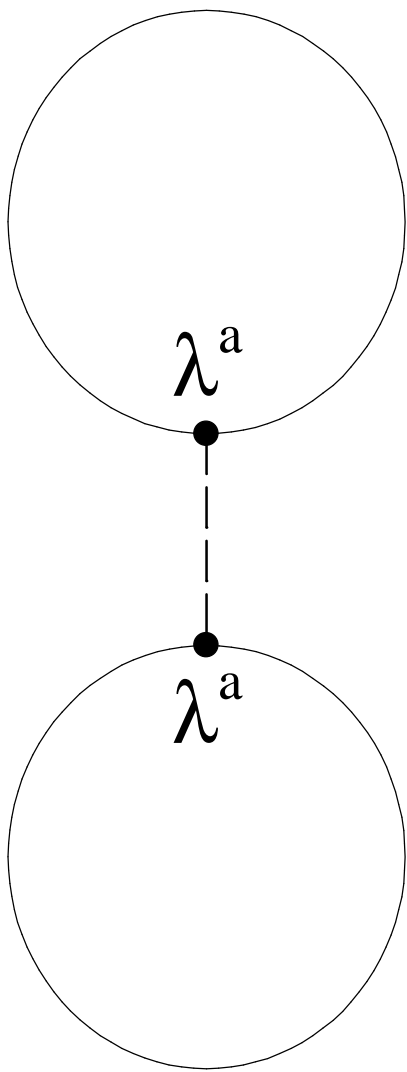
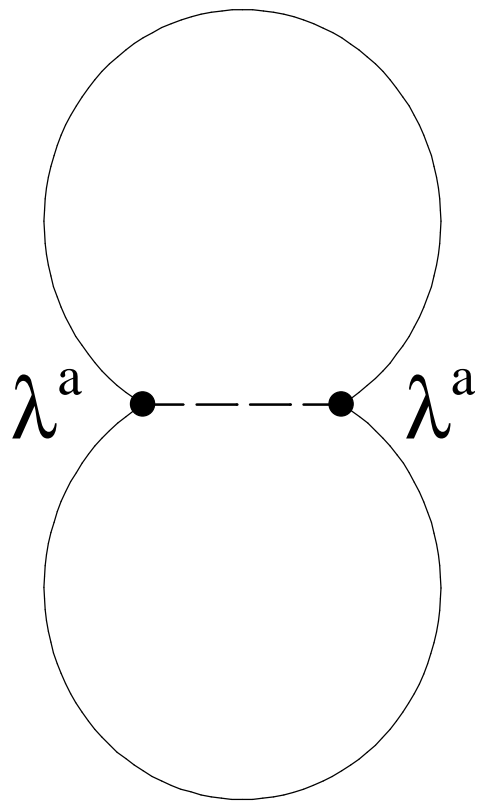
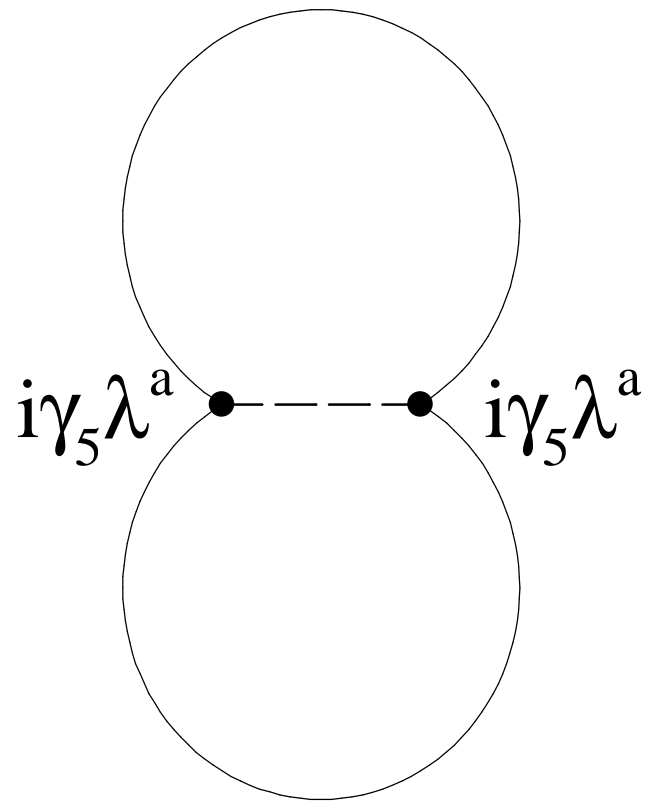
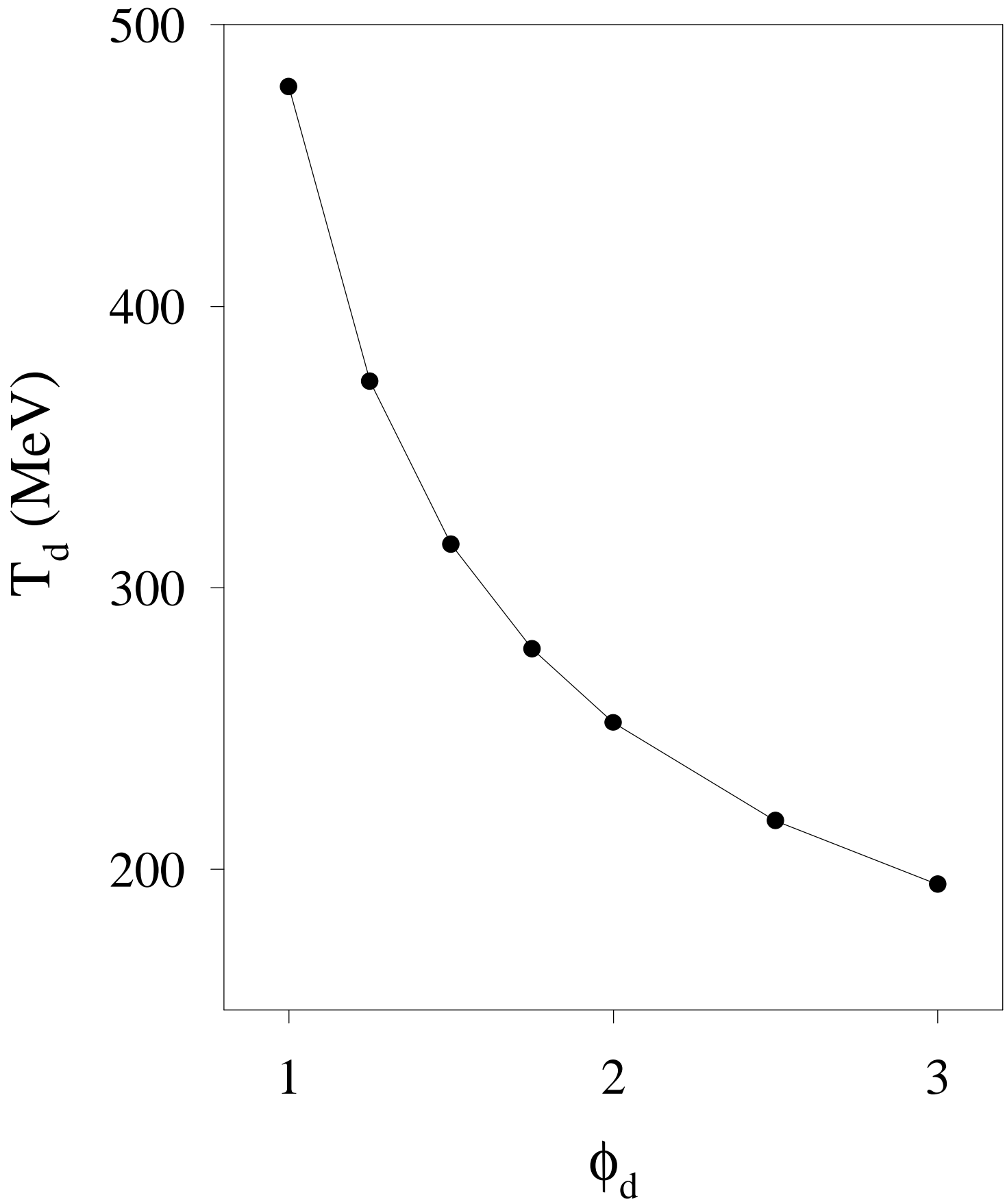
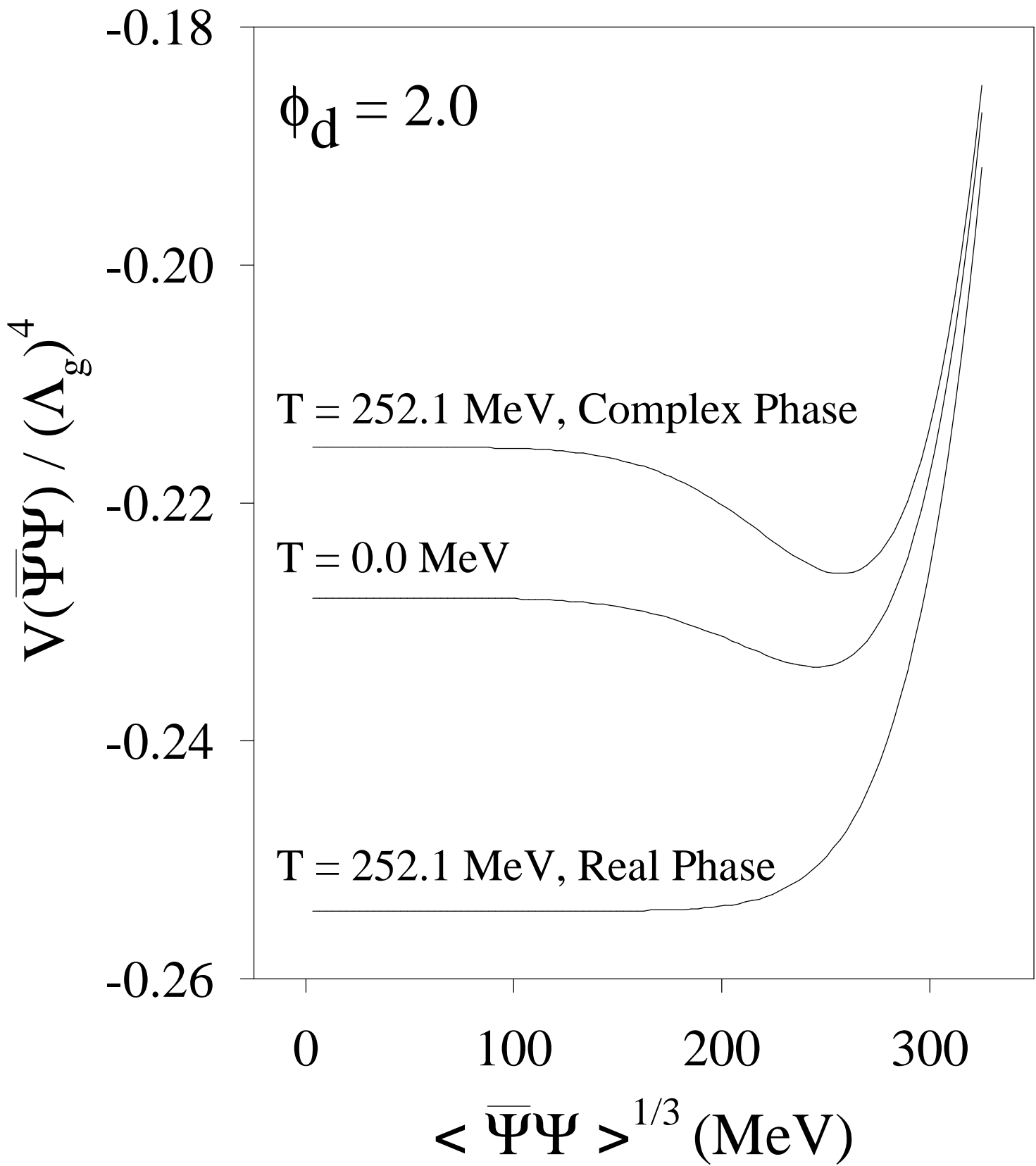
\end{document}